\documentclass[12pt]{article}
\usepackage{graphicx}
\usepackage{epsf}
\newcommand{\be}[1]{\begin{equation}\label{#1}}
\newcommand{\ee}{\end{equation}}
\newcommand{\bea}[1]{\begin{eqnarray}\label{#1}}
\newcommand{\eea}{\end{eqnarray}}
\newcommand{\micron}{\,\mu m}

\begin{document}
\vspace*{2cm}
\begin{center}
{\LARGE \bf The type of seeder cells determines\\
the efficiency of germinal center reactions}\\
\vspace{1cm}
Michael Meyer-Hermann and Tilo Beyer\\
\vspace{1cm}
Institut f\"ur Theoretische Physik, TU Dresden,
D-01062 Dresden, Germany\\
E-Mail: meyer-hermann@physik.tu-dresden.de\\
Correspondence should be addressed to MMH.
\end{center}

\vspace*{2cm}

\noindent{\bf Abstract:}
We discuss the origin of
two classes of germinal centers that have been observed 
during humoral immune responses: 
Some germinal centers develop very well and give rise to a 
large number of high affinity antibody producing plasma cells.
Other germinal center reaction are very weak and
the output production is practically absent. 
We propose an explanation for this nearly all-or-none behavior
of germinal center reactions: The affinity of the seeder
B-cells to the antigen is the critical parameter
that determines the fate of the germinal center
reaction. This hypothesis is verified in the framework
of a space-time simulation of germinal center
reactions.
\vspace*{\fill}
\eject
\newpage

\section{Introduction}

Germinal centers (GC) are an important part of the humoral
immune response \cite{Mac86,Ber91,Jac91,Nos91}. 
They generate new plasma cells that produce
antibodies of high affinity to a specific antigen. 
This affinity maturation process allows the organism to adapt its 
antibody repertoire to new antigens, to form the memory B cell
compartment, and
to react to antigens more efficiently compared to an immune
response using B-cells from the repertoire only.

Typically, GCs are found in lymph nodes or in the spleen. In reaction
to primary immunization some B-cells from the repertoire are
activated and become sensible to proliferation progression factors.
These B blasts (centroblasts) migrate towards the
primary follicle system. 
In the environment of follicular dendritic cells
and T cells they start a process of monoclonal expansion
\cite{Jac93,McH93,Pas94a}.
After about three days the centroblast population reaches about
12000 cells due to an extremely short cell cycle time of $6$-$7$
hours \cite{Han64,Zha88,Liu91}. 
During this early phase of the GC reaction a reduction of diversity
has been observed where those B cells of higher affinity to the
antigen have been preserved \cite{McH93}. This has been interpreted as
an early antigen-dependent preselection process
of un-mutated B cells.
The phase of monoclonal expansion is believed to be
followed by a phase of frequent somatic hypermutation \cite{Nos91,Ber87}, 
giving rise
to a great variety of different antibodies that are encoded in
the blasts. These cells start to differentiate to centrocytes
that are no longer in cell cycle but express large numbers of
antibodies \cite{Liu91,Cho00}. 
Centrocytes have initiated a process of apoptosis \cite{Liu94}.
They are rescued from apoptosis in dependence on an interaction
with antigens that are presented on the FDCs \cite{Liu89,Koo97,Rad98}. 
All centrocytes
that are not selected by such an interaction process die -- and
this is the large majority. Only positively selected centrocytes
get the chance to interact with T cells and, if again selected,
to further differentiate into antibody producing 
plasma cells or memory cells \cite{Pas94b,Smi97,Cho98,Sie01}.
As this selection process is believed to depend on the affinity
of antibodies and antigen, the GC reaction provides an
antibody optimization procedure.

GCs show some very interesting properties. Among those are
the separation of centroblasts and centrocytes into two zones,
the dark and the light zone \cite{Nos91,Liu91,Cam98}. 
The origin of those zones has been
discussed before in the framework of a stochastic space-time
model for the GC reaction \cite{Mey02}. Another property is
the nearly all-or-none behavior of GCs \cite{Rad98,Len01}. 
GC reactions of moderate
intensity have rarely been observed. The reaction is either
successful (concerning affinity maturation and number of produced
output cells) or strongly suppressed. The reason for this
behavior has not been resolved until now. Especially, the
all-or-none behavior could not be reproduced in a previous
modeling work \cite{Kle01}.

One suggestion
goes back to the starting point of the GC reaction,
the originally activated B cells \cite{Len01}. The authors
speculate that the mouse colony in their (pathogen-free) experiment may undergo
a {\it qualitatively and quantitatively diverse background immune
response}, i.e.~leading to a diversification of the B cell repertoire
between individuals. The consequence would be that the seeder cells
of the GC reaction would have different affinities to the antigen,
implying different GC reactions.
This leads us to the hypothesis that the quality
of GC seeder cells determines the fate of the GC reaction
\cite{Aga98,Vin00,Mey01}
and, in addition, may imply the all-or-none behavior of GCs.
In the present article this hypothesis is discussed 
in the framework of a previously
introduced stochastic space-time model for the GC reaction
\cite{Mey02}.

\section{Methods}

The model providing the basis of the analysis has
been introduced in all details before \cite{Mey02}. 
Therefore,
only those model elements are pointed out in details
that are of direct relevance in the present context. 
The basic concepts are only shortly summarized in the following.

Cells are represented on a fixed, equidistant, cubic, and
two dimensional lattice. Each cell occupies exactly one node
of the lattice. All cells can move on the lattice with some
velocity, which is anti-proportional to the cell radius. 
The movement is undirectioned and restricted by
the space available on the lattice. Each cell is of specific
type (centroblast or centrocyte) which determines their possible
actions: {\it Centroblasts} can divide. Each division allows a somatic
hypermutation to occur with probability $0.5$. They also can
differentiate to centrocytes provided that the differentiation process 
has been activated before. This happens if the centroblast binds a
certain amount of signal molecules, which is secerned by the
FDCs and diffuses over the lattice. Such a non-local interaction
process turned out to be crucial for the establishment of the
dark zone \cite{Mey02}. {\it Centrocytes} die by apoptosis and
are then eliminated from the lattice.
They may be rescued by an interaction with antigens on FDCs, which
becomes possible when centrocytes and FDCs are in direct contact
on the lattice and provided that the affinity of the presented antibodies
to the antigen is strong enough. 
Rescued centrocytes can further
differentiate to plasma- or memory-cells (output cells) or
restart to proliferate (i.e.~become centroblasts again).
{\it Output cells} leave the GC by diffusion. The parameters
that have been used here and that have been determined before
\cite{Mey02} are summarized in Tab.~\ref{parameter}.

Most important is the description of affinity between
antigen and antibodies. The model uses the traditional shape
space concept \cite{Per79}. The dimension of the shape space is
thought to represent phenotype properties of the antibodies.
The affinity to the antigen reaches a maximum at the shape
space position corresponding to optimal complementarity.
A mutation (only those of phenotypical relevance are counted)
is represented by a jump to a neighbor point in the shape space.
The principle of {\it affinity neighborhood} \cite{Mey01}
ensures that the affinity
may be increased by a sequence of somatic hypermutations. One
may think of climbing up an affinity hill. However, key mutations,
that has been frequently discussed (especially for immunization
with (4-hydroxy-3-nitrophenyl)acetyl (NP) \cite{Rad98})
are not easily covered by this concept. Therefore, the outcome
of the analysis will have to be interpreted in this context.

The quality of a B cell is defined by its distance to the
antigen (i.e. to the optimal B cell clone). The distance is
calculated by the number of (phenotypically relevant) somatic
hypermutations that are necessary to reach the antigen position
in the shape space. The number of mutations is
denoted by {\it mutation-distance} in the
following. This corresponds to the definition of a
1-norm on the shape space which is assumed to have four dimensions.
Using this norm the affinity of antibody and antigen is quantitatively
described by a Gaussian function which is centered at the antigen
position $\Phi_0$ in the shape space \cite{Mey01}. 
In the following this concept is used to describe the {\it efficiency}
of the GC reaction in a qualitative and quantitative way.
At first, the quality of a GC reaction is measured by the
success of the affinity maturation process. To this end
the fraction of high affinity output cells (the sum of plasma
and memory cells) resulting from the GC reaction is analyzed.
{\it High affinity} means that the antibody binds the antigen
with a probability of more than $30\%$. Secondly, the number
of resulting output cells (summed over the whole GC reaction)
is a good measure for the quantitative efficiency of the 
GC reaction.

\begin{figure}[ht]
\begin{center}
\includegraphics[height=7.5cm]{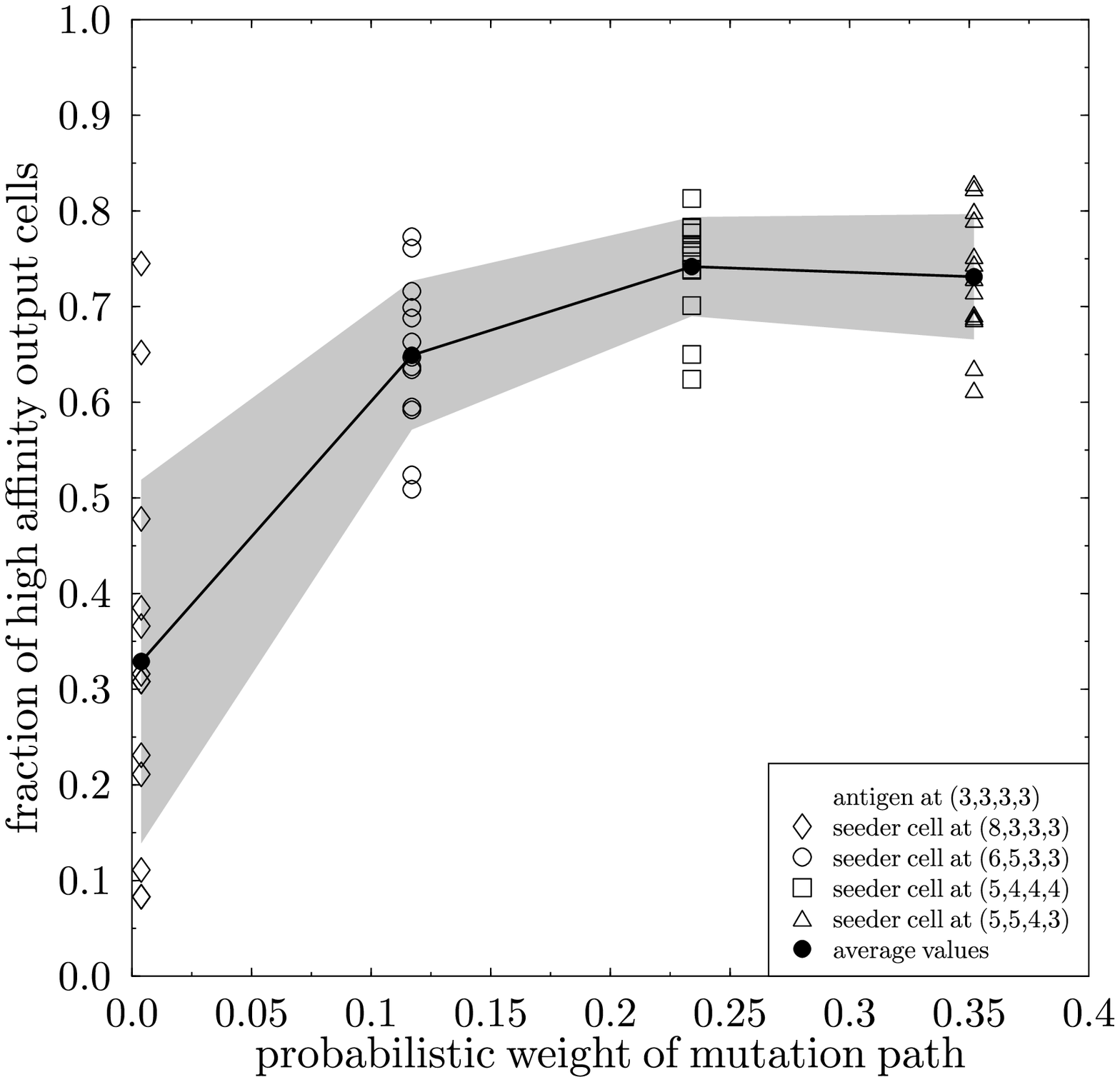}
\includegraphics[height=7.5cm]{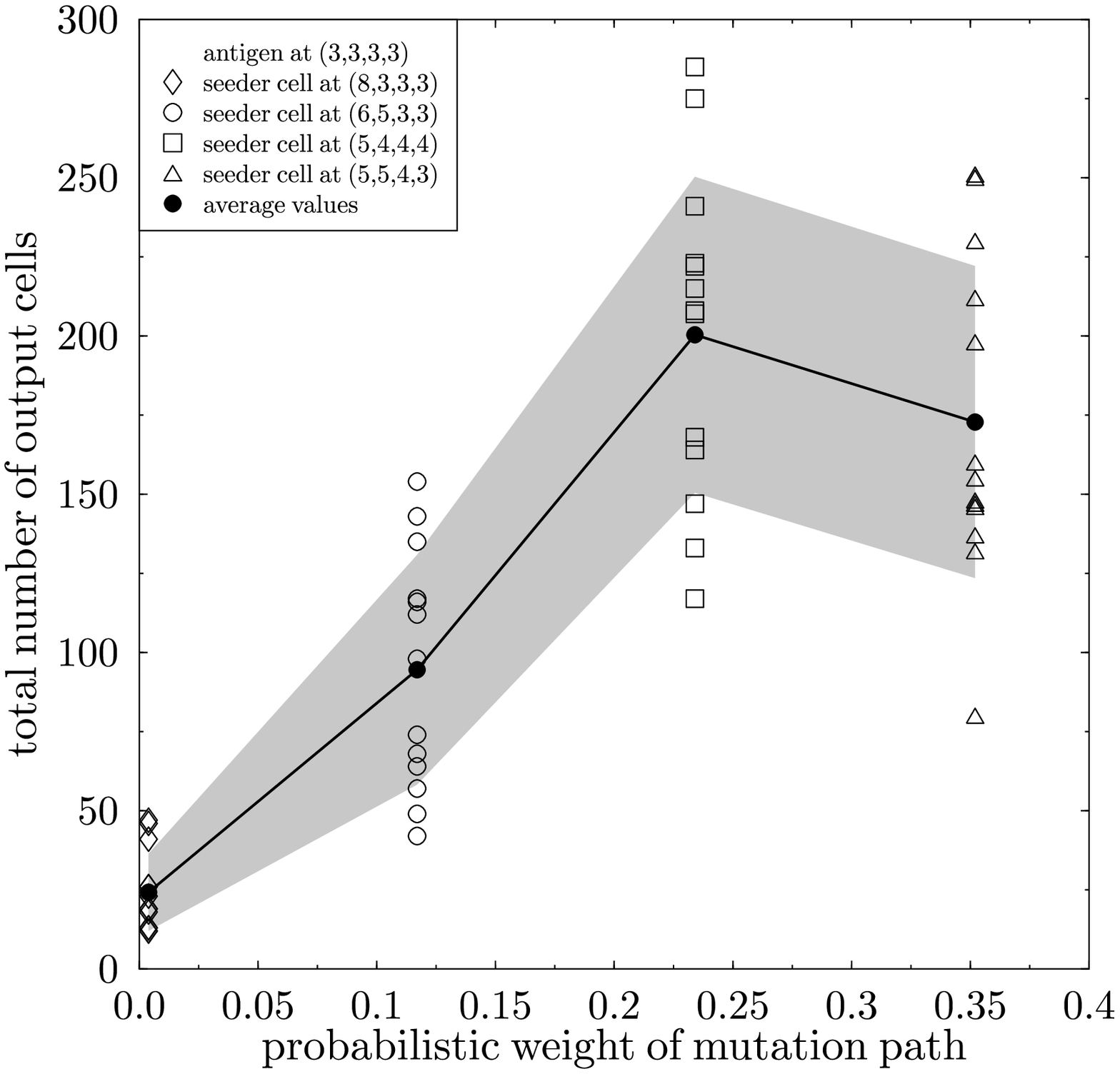}
\end{center}
\vspace*{-5mm}
\caption[]{\sf The fraction of high affinity output cells and the 
number of output cells resulting from GC reactions
for seeder cells with a distance of 5 mutations to
the optimal clone at position $(3,3,3,3)$ in the shape space but
for different stochastic weights $p$ of the corresponding mutation paths.
The seeder cells are at position 
$(8,3,3,3)$ (diamonds, $p=0.0039$), 
$(6,5,3,3)$ (circles, $p=0.12$),
$(5,4,4,4)$ (squares, $p=0.23$),
and $(5,5,4,3)$ (triangles, $p=0.35$).
The corresponding average values from $13$ GC reactions (filled circles) 
are shown with one standard deviation (gray area).
The output cell quality grows and saturates for higher stochastic
weights, which is less clear for the number of output cells.}
\label{seed_5n}
\end{figure}

However, the above notion of B cell quality hides an intrinsic
ambiguity. Let the antigen be at position $\Phi_0=(3,3,3,3)$ and
let the seeder cells have a mutation-distance of $5$.
The outcome of the GC reaction depends on the
exact position of the seeder cell, which may have the
initial position $\Phi_1=(8,3,3,3)$ or $\Phi_2=(6,5,4,4)$. Assuming 
the mutations to occur randomly \cite{Rad98,Wei70},
the probabilistic
weights $p(\Phi)$ of both seeder cells to find the antigen
position $\Phi_0$ considerably differs: $p(\Phi_1)=1/256$ and
$p(\Phi_2)=90/256$. The difference is due to the number of possible
permutations in the sequence of mutations.
In other words, there exist more combinatorial 
possibilities for the mutations
that all lead to the same target $\Phi_0$.
This ambiguity is mirrored in the GC efficiency 
(quality and quantity of output cells, see Fig.~\ref{seed_5n}).
Less probable mutation paths induce GC reactions of smaller
efficiency. However, at least the fraction of high affinity
output cells saturates for more probable mutation paths.
In order to ensure a well-defined mutation distance, only
seeder cells that reach the antigen on a mutation path of
maximum probability will be taken into account. The results
have been cross-checked using mutation paths of moderate probability.

Using the thus defined model concepts we initiate GC reactions
with various types of seeder cells. The seeder cells are placed at random
positions in the FDC network. The mutation-distance is varied between
$0$ and $12$. Mutation-distances to the antigen of more than $12$
seem to be unphysiological \cite{Kue93,Wed97}.
Only a cutout of the shape space in the neighborhood of the antigen
(which is fixed at $\Phi_0$ in the shape space) is considered: The range
of each shape space dimension is $0 \le \Phi_i \le 10$. B cells that
leave this shape space area are considered to have vanishing affinity
to the antigen. The positions
of the seeder cells are listed in Tab.~\ref{seedertypes}.
As we use a stochastic model, the GC reactions are repeated with
different initializations of the random number generator. Each result
is presented by showing $13$ GC reactions, the average value, and
one standard deviation.

\section{Results}

We find a clear correlation of the total
number of output cells produced during the GC reactions and
the mutation-distance of the seeder cells that initiate the
GC reaction (see Fig.~\ref{seed2_on}).
If the seeder cells are already high affinity B cells, the
GC reaction is rather stable and produces a large amount
of high affinity output cells. With increasing mutation-distance
the number of output cells is reduced in a sigmoidal-like
functional dependence. For mutation-distances of more than
$6$ the outcome of the GC reaction becomes critical and for more
than $9$ the GC reaction is not able to generate a relevant
amount of output cells.

\begin{figure}[ht]
\begin{center}
\includegraphics[height=11cm]{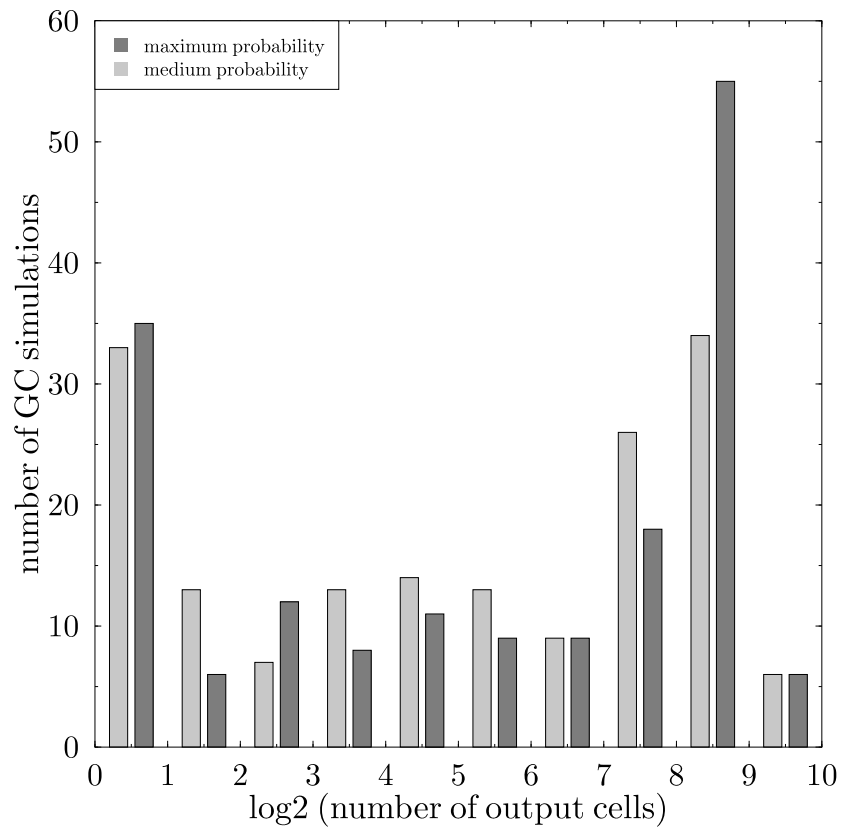}
\end{center}
\vspace*{-5mm}
\caption[]{\sf The number of GC simulations for different numbers
of output cells (the exponent of $2$ is shown on the abscissae).
The seeder cells are at a mutation-distance in the range
of $0$ and $12$. The data are based on 26 simulations for
each mutation-distance ($=338$ simulations).
The result is shown for mutation paths of maximum (dark gray)
and moderate (light gray) probability, separately.
The all-or-none behavior of GC reactions is clearly
reproduced.}
\label{seed2_dn}
\end{figure}

The GC reaction breaks down in a relatively narrow range of
the seeder cell mutation-distance to the antigen. 
This points towards an all-or-none
development of GC reactions which depends on the seeder
cell quality (at least concerning the output quantity).
This interpretation is verified by counting the numbers
of GC reactions that lead to the production of different
numbers of output cells (see Fig.~\ref{seed2_dn}). 
This analysis
is based on the same number of GC reactions for each
type of seeder cells. Indeed,
GCs preferentially produce large or nearly vanishing
amounts of output cells. Small or moderate amounts are
rare events. 

Note, that these results remain basically unaltered
considering seeder cells with mutation paths to the antigen
of moderate (instead of maximum) probabilistic
weight (see Tab.~\ref{seedertypes}, Fig.~\ref{seed2_on}
and \ref{seed2_dn}). However, the break-down of the GC
reaction is slightly shifted to seeder cells of higher quality.

\begin{figure}[ht]
\begin{center}
\includegraphics[height=11cm]{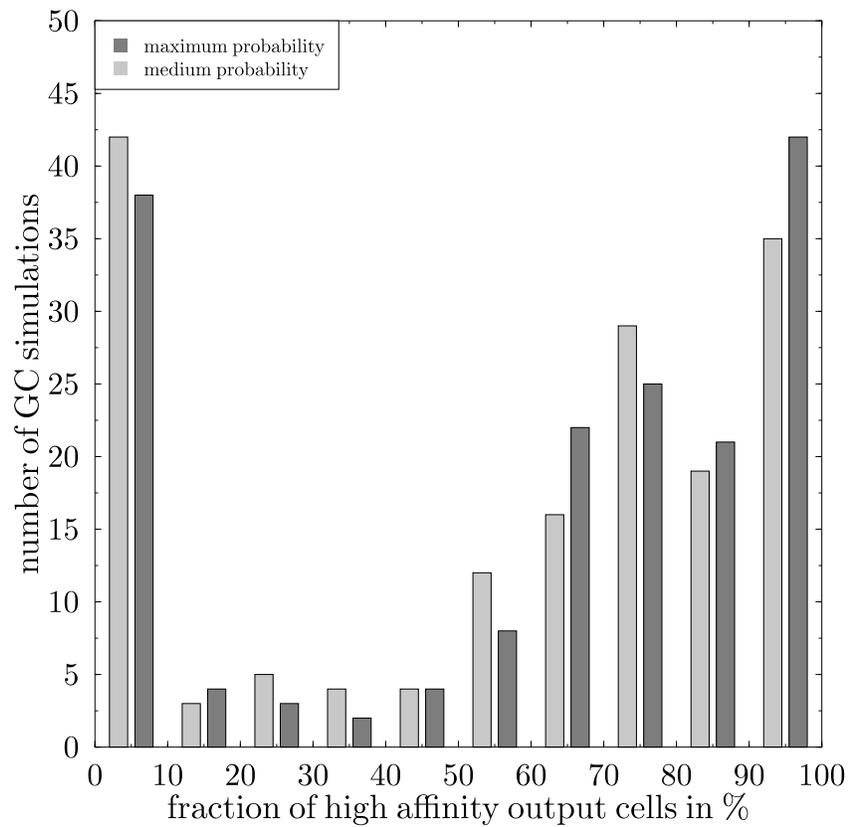}
\end{center}
\vspace*{-5mm}
\caption[]{\sf The number of GC simulations 
leading to different fractions of high affinity output cells (in \%).
The seeder cells are at a mutation-distance in the range
of $0$ and $12$. The data are based on $26$ simulations for
each mutation-distance ($=338$ simulations).
The result is shown for mutation paths of maximum (dark gray)
and moderate (light gray) probability, separately.
Again, the all-or-none behavior of GC reactions is clearly
reproduced.}
\label{seed2_dq}
\end{figure}

The output quality is correlated to the seeder cell quality
in a very analogous way (see Fig.~\ref{seed2_oq}). 
The GC reaction produces a stable and large fraction of 
high affinity output cells for high affinity seeder cells.
For seeder cells with larger mutation-distances 
the fraction of high affinity output decreases.
The correlation becomes instable for very large
mutation-distances. This is due
to the small number of produced output cells
in this regime. Large statistical fluctuations
obscure the correlation of output and seeder cell
quality. However, plotting the number of GC reaction
leading to some fraction of high affinity output cells
(see Fig.~\ref{seed2_dq}) shows that
the all-or-none behavior of GCs is clearly found for the
output quality as well. Note, that this result is
again not affected by switching to seeder cells with
mutation paths to the antigen of
moderate (instead of maximum) probabilistic weight.

An interesting observation of the GC reactions for
various types of seeder cells is that the dark zone
is established in all GC reaction in a very similar
way -- independently of the seeder cell quality.
The dark zone is always depleted around day $8$ 
of the reaction (data not shown).
This implies that the initial phase of the GC reaction
is not influenced by the seeder cell affinity to the
antigen. This initial phase includes the phase of
monoclonal expansion and even the first days of the
phase of somatic hypermutation and selection (what we
have called the primary optimization phase \cite{Mey01}). 
It is only afterwards when the selection of centrocytes
becomes the dominant process of the GC reaction, that the
seeder cell affinity to the antigen is a critical parameter.
We observe a huge sensitivity of the total GC life time
on the seeder cell quality, which varies from $12$
to $27$ days. This implies that if the B cell quality
is not substantially enhanced during the first phase
of the reaction, the GC B cell population rapidly dies
out within $4$ to $5$ days --
a behavior that has been observed in experiment as well
\cite{Han95,Vin00}. 

It is important to verify 
that the affinity maturation process was, indeed, unsuccessful 
in rapidly dying GCs.
This is best shown by
considering the number of recycled centroblasts (i.e.~re-proliferating
centroblasts that have been positively selected at least once)
after $10$ days (a representative time point) of the reaction. 
If a centrocyte is positively selected it will restart to
proliferate with a high probability \cite{Mey01} and therefore
the number of recycled cells is a good measure for the
success of the affinity maturation process.
For seeder cells with
a mutation-distance of more than $8$ this value remains
below $20$ and reaches $150$ for seeder cells with mutation-distance
of $3$.

Let us have a deeper look at the regime where the GC reaction
becomes critical. This is the case for seeder cells with
mutation-distances from $6$ and $7$.
The same analysis as performed in Figs.~\ref{seed2_dn} and
\ref{seed2_dq} is repeated taking into account the corresponding
GC simulations only. The result is shown in
Fig.~\ref{see2_dqnx}. The all-or-none behavior seen before
has disappeared. The distribution of GCs with respect to
the fraction of high affinity output cells is peaked at
$65\,\%$ and flat for smaller fractions. In the case
of the number of output cells the distribution
of GC reactions is clearly peaked at about 50 cells
in a Gaussian-like shape. This is the result expected
if the all-or-none behavior of GCs is indeed correlated
to the seeder cell quality. If, considering a small
window of mutation-distances for the seeder cells,
the all-or-none behavior
still appeared as it did in Figs.~\ref{seed2_dn} and
\ref{seed2_dq}, the reason for the all-or-none behavior
would have to be looked for somewhere else in the GC
dynamics.

\section{Discussion}

The present analysis has revealed that the seeder cell quality
(measured as the mutation-distance to the optimal clone) determines
the fate of the GC reaction and is responsible for the all-or-none
development of GC reactions as observed in experiments.
The efficiency of the GC reaction has been measured quantitatively
and qualitatively considering number and affinity of the produced
output cells. This result implies that the repertoire of B cells
in an organism determines whether the immune system may respond with
GC reactions in an efficient way or not. 
If the repertoire is reduced under some threshold, 
the probability of a successful GC reaction may become
critically small for some specific antigens. 

\begin{figure}[ht]
\begin{center}
\includegraphics[height=11cm]{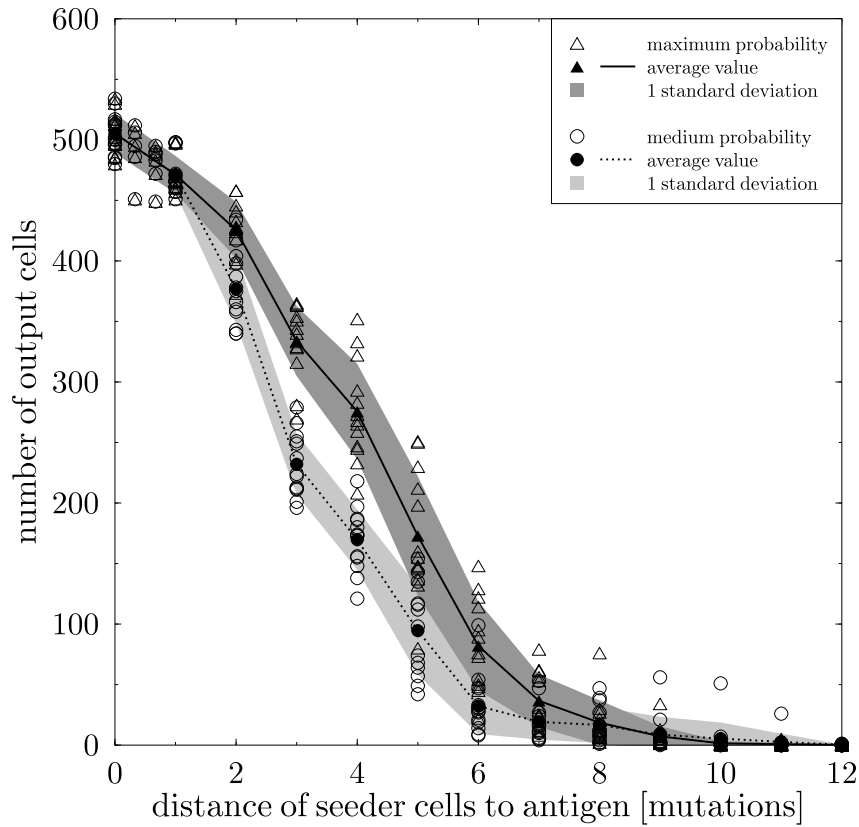}
\end{center}
\vspace*{-5mm}
\caption[]{\sf The number of output cells resulting from $338$ GC
reactions for different seeder cell qualities. The results
are shown for mutation paths of maximum ($169$ triangles)
and moderate ($169$ circles) probability. Filled triangles and
circles denote the corresponding average values, respectively.
The dark and light gray area denotes one standard deviation
for each type of mutation path.
The number of output cells grows with the seeder cell
quality in a sigmoidal-like way. More probable mutation
paths shift the sigmoidal function to seeder cells of lower
quality.}
\label{seed2_on}
\end{figure}

\begin{figure}[ht]
\begin{center}
\includegraphics[height=11cm]{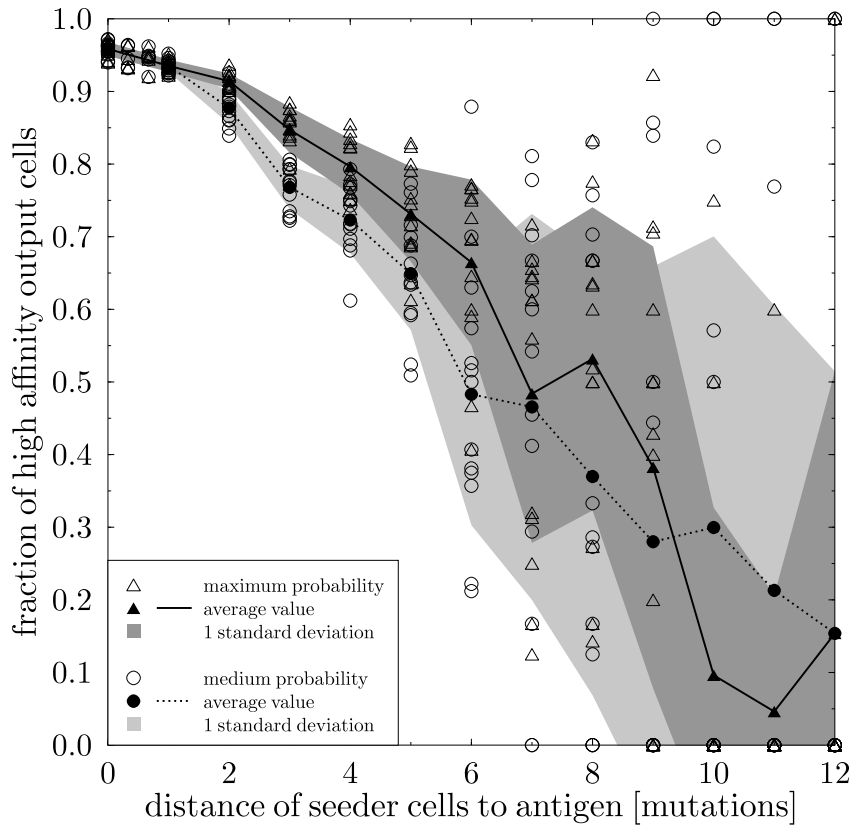}
\end{center}
\vspace*{-5mm}
\caption[]{\sf The fraction of high affinity output cells resulting in 
$338$ GC reactions for different seeder cell qualities. The results
are shown for mutation paths of maximum (triangles)
and moderate (circles) probability. Filled triangles and
circles denote the corresponding average values, respectively.
The dark and light gray area denotes one standard deviation
for each type of mutation path.
The output cell quality decreases for lower seeder cell
quality. For bad quality seeder cells the GC reaction becomes
unstable. For the most probable mutation
path the seeder cell quality necessary for a successful GC reaction
is slightly lowered.}
\label{seed2_oq}
\end{figure}

However, there are some model assumptions that should be
revisited. At first, the model is two-dimensional.
In three dimensions the number of B cells in a fully developed
GC is substantially larger.
Therefore, we expect a shift of the curves in Fig.~\ref{seed2_on}
and \ref{seed2_oq} to higher mutation-distances of the
seeder cells. This shift is nontrivial and
difficult to be estimated quantitatively because the translation
from two to three dimension may also change
the interaction frequency of centrocytes and FDCs.

In view of the shape space concept a possible role of key
mutations may affect the results even when the key mutations
are in a cold spot as in the case of NP \cite{Rad98}. However,
one may expect that the existence of key mutations would
even intensify the all-or-none behavior of GCs because
the B cells are confronted with an additional critical
process, namely to find the key mutation or not. 
Therefore, one may suspect that the sigmoidal function
in Fig.~\ref{seed2_on} may even become steeper. 
From this perspective the present
results show that the existence of key mutations is not
necessary to understand the all-or-none development of
GC reactions \cite{Rad98}. 
The latter follows already within a 
quasi-continuous affinity maturation concept.

\begin{figure}[ht]
\begin{center}
\includegraphics[height=7.5cm]{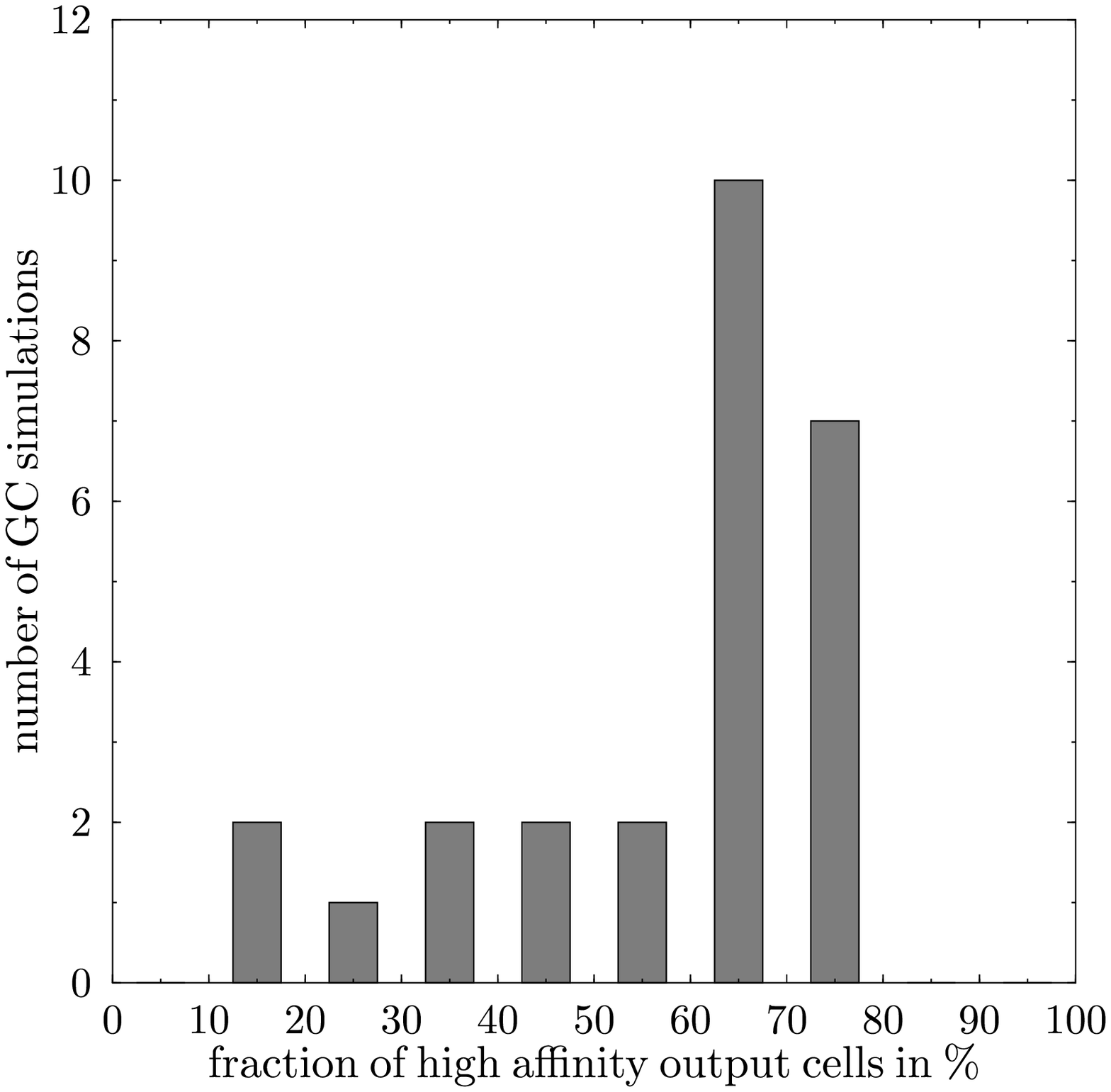}
\includegraphics[height=7.5cm]{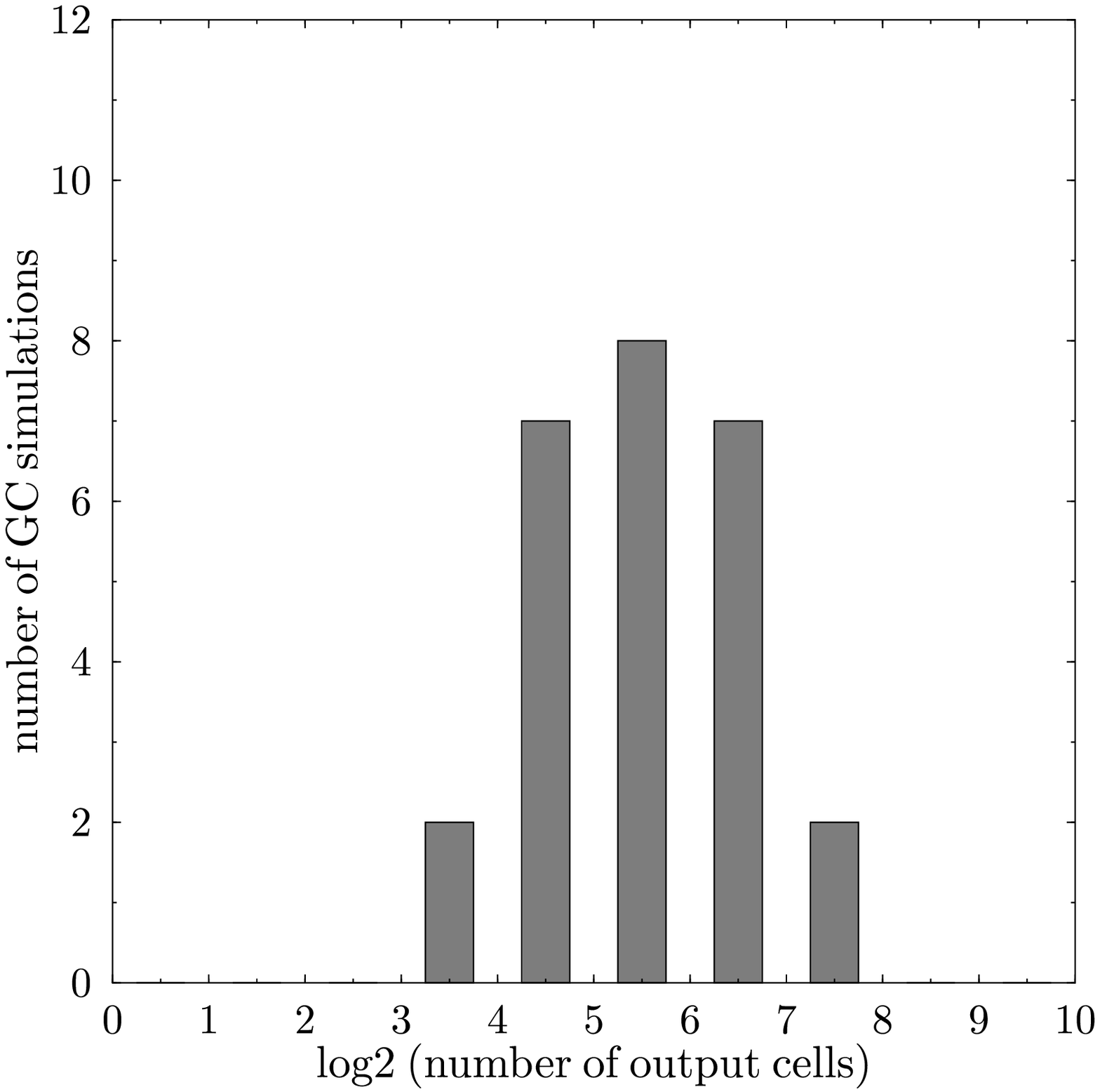}
\end{center}
\vspace*{-5mm}
\caption[]{\sf The number of GC simulations 
leading to different fractions of high affinity output cells (in \%)
and to different numbers of output cells 
(the exponent of $2$ is shown on the abscissae).
Only those simulations are shown that started with
seeder cells in a critical distance to the optimal mutant
($6$ and $7$). The data are based on 13 simulations for
each mutation-distance ($=26$ simulations).
The result is shown for mutation paths of maximum probability.
The all-or-none behavior of GC reactions found in Fig.~\ref{seed2_dq}
and \ref{seed2_dn} disappears. The distribution of GCs
with respect to the number of output cells even 
becomes Gaussian-like.}
\label{see2_dqnx}
\end{figure}

The analysis is based on the assumption that the seeder
cells are homogeneously distributed with respect
to the mutation-distances to the antigen. This assumption
is not essential for the outcome provided that not all
activated B cells that give rise to a GC reaction always
have the same mutation-distance to the antigen
(see Fig.~\ref{see2_dqnx}). It is reasonable to expect
that the mutation-distance of activated B cells is
basically a random parameter. 
Indeed, it has recently been found for T cell-dependent primary
immune responses to NP, that high and low
affinity B lymphocytes are equally suitable to induce
GC reactions \cite{Shi02}. The affinity of B cells to
the antigen differed by a factor of 40 in those measurements.
Therefore, we expect that deviations
from our assumption of homogeneity will remain sufficiently
small not to affect our interpretation of the results.

An important outcome of the analysis is that the seeder cell
quality does not affect the initial phase of the GC reaction
(compare also \cite{Shi02}).
One may have suspected that the GC reaction simply doesn't
start if the seeder cells have subthreshold affinity to the
antigen \cite{Aga98}
(it is assumed that the affinity is sufficient
to activate some naive B cells). 
This possibility is not excluded by the present
analysis, as there may be some process, which is not included
in the model, but which is crucial for the initiation of the GC
reaction. However, our computer experiment shows that the
all-or-none behavior can be understood on the level of 
the efficiency of GC reactions without postulating an additional
process during the initiation of the GC reaction. 

This result is in accordance with observations of T cell independent
GC reactions \cite{Len01,Vin00}. Here, it has been found that GCs
develop normally in absence of T cell help. Even the dark and
light zones are established \cite{Vin00}. 
As T cells are believed
to be of great importance for the selection process (especially
to inhibit apoptosis of centrocytes \cite{Cho96,Hol99,Eij01}) 
the author claimed
that the initial phase is basically independent of T cell help.
In the present context, this implies that provided that B cells have
been activated by the antigen, they may induce a GC reaction
independently of their affinity to the antigen (note that
assuming a preselection process of un-mutated B cells \cite{McH93}, 
this statement applies to positively pre-selected B cells). 
In a later
phase of the GC reaction the T cell help is crucial for
centrocytes to survive and, consequently, the already well
established GCs are rapidly depleted without T cells
\cite{Len01,Vin00}. The same happens in the present simulations
but for a different reason: While in T cell independent GC
reactions the selection process cannot start at all, bad
quality seeder cells may still allow for a weak selection
process, which is, however, insufficient to achieve a successful 
affinity maturation process. In both scenarios -- in T cell
independent GCs and for GCs initiated by low quality seeder cells 
-- the GCs die out because of a failing selection procedure.

The differences of GC responses
to two similar (but non-identical) peptides (PS1CT3 and G32CT3)
have been examined \cite{Aga98}.
The authors found that the intensity of a GC reaction
is dependent on the used peptide and, consequently, dependent on
the quality of the initially activated B cells.
The results found here support these findings, and, in addition,
specify the character of this dependence:
The seeder cell quality induces an all-or-none behavior of
the GC reaction intensity.





%
%
%

\newpage


%
\begin{table}[ht]
\small
\begin{center}
\begin{tabular}{|l|c|c|} \hline
Parameter & symbol & value\\ \hline
Shape space dimension                    &$d_s$&$4$\\
Width of Gaussian affinity weight function &$\Gamma$&$2.8$\\
Lattice constant                         &$\Delta x$&$10\micron$\\
Radius of GC                             &$R$&$220\micron$\\
Number of seeder cells                   &$$&$3$\\
Diffusion constant for centroblasts      &$D_{CB}$&$5\frac{\micron^2}{hr}$\\
Ratio of centroblast to centrocyte radius&$\frac{r_{CB}}{r_{CC}}$&$3$\\
Diffusion constant of signal molecules   &$D_{\rm sig}$&$200\frac{\micron^2}{hr}$\\
Number of FDCs                           &$$&$20$\\
Length of FDC arms                       &$$&$30\micron$\\
Duration of phase of monoclonal expansion&$\Delta t_1$&$72\,hr$\\
Duration of optimization phase           &$\Delta t_2$&$48\,hr$\\
Rate of proliferation (2D)               &$p$&$1/(9hr)$\\
Maximal distance for CB proliferation    &$R_{\rm P}$&$60\micron$\\
Mutation probability                     &$m$&$0.5$\\
Rate of differentiation signal production by FDCs&$s$&$9/hr$\\
Rate of centroblast differentiation      &$g_1$&$1/(6hr)$\\
Rate of FDC-centrocyte dissociation      &$g_2$&$1/2hr$\\
Rate of differentiation of selected centrocytes&$g_3$&$1/(7hr)$\\
Probability of recycling of selected centrocytes&$q$&$0.8$\\ 
Rate of centrocyte apoptosis             &$z$&$1/(7hr)$\\
\hline
\end{tabular}
\caption[]{\sf Summary of all model parameters. The values have
been determined using corresponding experimental data \cite{Mey02}.
Note, that all given rates are the physiological
ones which enter with an additional factor 
of $\ln(2)$ into the model. The symbols correspond
to those used in the text.}
\label{parameter}
\end{center}
\end{table}
\begin{table}[ht]
\small
\begin{center}
\begin{tabular}{|r|c|c|c|l|} \hline
mutation-distance & \multicolumn{3}{c}{position} \vline
& mutation path weight\\
$||\Phi-\Phi_0||_1$
& \multicolumn{1}{c}{$\Phi_1$}
& \multicolumn{1}{c}{$\Phi_2$}
& \multicolumn{1}{c}{$\Phi_3$} \vline 
& \\ \hline
0 & 3333 & 3333 & 3333 & maximum\\
1 & 3334 & 3343 & 3433 & maximum\\
2 & 3335 & 3353 & 3533 & moderate\\
2 & 3344 & 3443 & 4433 & maximum\\
3 & 3336 & 3363 & 3633 & moderate\\
3 & 3345 & 3453 & 4533 & maximum\\
4 & 3346 & 3463 & 4633 & moderate\\
4 & 3445 & 4453 & 4534 & maximum\\ \hline
5 & 3338 & 3383 & 3833 & low\\
5 & 3356 & 3563 & 5633 & moderate\\
5 & 4445 & 4454 & 4544 & high\\
5 & 3455 & 4553 & 5534 & maximum\\ \hline
6 & 3366 & 3663 & 6633 & moderate\\
6 & 6543 & 5436 & 4365 & maximum\\
7 & 3466 & 4663 & 6634 & moderate\\
7 & 4456 & 4564 & 5644 & maximum\\
8 & 3566 & 5663 & 6635 & moderate\\
8 & 6554 & 5546 & 5465 & maximum\\
9 & 3666 & 6663 & 6636 & moderate\\
9 & 6654 & 6546 & 5466 & maximum\\
10& 4666 & 6664 & 6646 & moderate\\
10& 7654 & 6547 & 5476 & maximum\\
11& 5666 & 6665 & 6656 & moderate\\
11& 7655 & 6557 & 5576 & maximum\\
12& 5667 & 6675 & 6756 & maximum\\
\hline
\end{tabular}
\caption[]{\sf The positions $\Phi_1$, $\Phi_2$, and $\Phi_3$ in the shape space
of the three seeder cells that initiate the GC reaction. The antigen is at
position $\Phi_0=(3333)$. The mutation-distance is varied between 
$0$ and $12$. The mutation-paths are grouped according to their
probabilistic weight defining sets of used seeder cells.}
\label{seedertypes}
\end{center}
\end{table}
%
%
%
%


\begin{thebibliography}{99}


\bibitem{Mac86}
\normalsize
{\sc MacLennan, I.\,/\,Gray, D.},
\normalsize
{\rm Antigen-driven selection of virgin and memory B cells}.
\normalsize
{\it Immunol. Rev.\/}
\normalsize
{\bf 91}
\normalsize
{\rm (1986)},
\normalsize
{\rm 61}.
\normalsize

\bibitem{Ber91}
\normalsize
{\sc Berek, C.\,/\,Berger, A.\,/\,Apel, M.},
\normalsize
{\rm Maturation of the immune response in the germinal center}.
\normalsize
{\it Cell\/}
\normalsize
{\bf 67}
\normalsize
{\rm (1991)},
\normalsize
{\rm 1121-1129}.
\normalsize

\bibitem{Jac91}
\normalsize
{\sc Jacob, J.\,/\,Kelsoe, G.\,/\,Rajewski, K.\,/\,Weiss, U.},
\normalsize
{\rm Intraclonal generation of antibody mutants in germinal 
centres}.
\normalsize
{\it Nature\/}
\normalsize
{\bf 354}
\normalsize
{\rm (1991)},
\normalsize
{\rm 389}.
\normalsize

\bibitem{Nos91}
\normalsize
{\sc Nossal, G.},
\normalsize
{\rm The molecular and cellular basis of affinity maturation 
in the antibody response}.
\normalsize
{\it Cell\/}
\normalsize
{\bf 68}
\normalsize
{\rm (1991)},
\normalsize
{\rm 1-2}.
\normalsize

\bibitem{Jac93}
\normalsize
{\sc Jacob, J.\,/\,Przylepa, J.\,/\,Miller, C.\,/\,Kelsoe, G.},
\normalsize
{\rm In situ studies of the primary response to (4-hydroxy-3-nitrophenyl)acetyl. 
III. The kinetics of V region mutation and selection 
in germinal center B cells}.
\normalsize
{\it J. Exp. Med.\/}
\normalsize
{\bf 178}
\normalsize
{\rm (1993)},
\normalsize
{\rm 1293-1307}.
\normalsize

\bibitem{McH93}
\normalsize
{\sc McHeyzer-Williams, M.G.\,/\,McLean, M.J.\,/\,Labor, P.A.\,/\,Nossal, 
G.V.J.},
\normalsize
{\rm Antigen-driven B cell differentiation in vivo}.
\normalsize
{\it J. Exp. Med.\/}
\normalsize
{\bf 178}
\normalsize
{\rm (1993)},
\normalsize
{\rm 295-307}.
\normalsize

\bibitem{Pas94a}
\normalsize
{\sc Pascual, V.\,/\,Liu, Y.-J.\,/\,Magalski, A.\,/\,De Bouteiller, 
O.\,/\,Banchereau, J.\,/\,Capra, J.D.},
\normalsize
{\rm Analysis of somatic mutation in five B cell subsets 
of human tonsil}.
\normalsize
{\it J. Exp. Med.\/}
\normalsize
{\bf 180}
\normalsize
{\rm 1994},
\normalsize
{\rm 329-339}.
\normalsize

\bibitem{Han64}
\normalsize
{\sc Hanna, M.G.},
\normalsize
{\rm An autoradiographic study of the germinal center in 
spleen white pulp during early intervals of the immune 
response}.
\normalsize
{\it Lab. Invest.\/}
\normalsize
{\bf 13}
\normalsize
{\rm (1964)},
\normalsize
{\rm 95-104}.
\normalsize

\bibitem{Zha88}
\normalsize
{\sc Zhang, J.\,/\,MacLennan, I.C.M.\,/\,Liu, Y.J.\,/\,Land, P.J.L.},
\normalsize
{\rm Is rapid proliferation in B centroblasts linked to 
somatic mutation in memory B cell clones}.
\normalsize
{\it Immunol. Lett.\/}
\normalsize
{\bf 18}
\normalsize
{\rm (1988)},
\normalsize
{\rm 297-299}.
\normalsize

\bibitem{Liu91}
\normalsize
{\sc Liu, Y.J.\,/\,Zhang, J.\,/\,Lane, P.J.\,/\,Chan, E.Y.\,/\,MacLennan, 
I.C.M.},
\normalsize
{\rm Sites of specific B cell activation in primary and 
secondary responses to T cell-dependent and T cell-independent 
antigens}.
\normalsize
{\it Eur. J. Immunol.\/}
\normalsize
{\bf 21}
\normalsize
{\rm (1991)},
\normalsize
{\rm 2951-2962}.
\normalsize

\bibitem{Ber87}
\normalsize
{\sc Berek, C.\,/\,Milstein, C.},
\normalsize
{\rm Mutation drift and repertoire shift in the maturation 
of the immune response}.
\normalsize
{\it Immunol. Rev.\/}
\normalsize
{\bf 96}
\normalsize
{\rm (1987)},
\normalsize
{\rm 23-41}.
\normalsize

\bibitem{Cho00}
\normalsize
{\sc Choe, J.\,/\,Li, L.\,/\,Zhang, X.\,/\,Gregory, C.D.\,/\,Choi, Y.S.},
\normalsize
{\rm Distinct Role of Follicular Dendritic Cells and T Cells 
in the Proliferation, Differentiation, and Apoptosis 
of a Centroblast Cell Line, L3055}.
\normalsize
{\it J. Immunol.\/}
\normalsize
{\bf 164}
\normalsize
{\rm (2000)},
\normalsize
{\rm 56-63}.
\normalsize

\bibitem{Liu94}
\normalsize
{\sc Liu, Y.J.\,/\,Barthelemy, C.\,/\,De Bouteiller, O.\,/\,Banchereau, 
J.},
\normalsize
{\rm The differences in survival and phenotype between centroblasts 
and centrocytes}.
\normalsize
{\it Adv. Exp. Med. Biol.\/}
\normalsize
{\bf 355}
\normalsize
{\rm (1994)},
\normalsize
{\rm 213-218}.
\normalsize

\bibitem{Liu89}
\normalsize
{\sc Liu, Y.J.\,/\,Joshua, D.E.\,/\,Williams, G.T.\,/\,Smith, C.A.\,/\,Gordon, 
J.\,/\,MacLennan, I.C.},
\normalsize
{\rm Mechanism of antigen-driven selection in germinal centres}.
\normalsize
{\it Nature\/}
\normalsize
{\bf 342}
\normalsize
{\rm (1989)},
\normalsize
{\rm 929-931}.
\normalsize

\bibitem{Koo97}
\normalsize
{\sc Koopman, G.\,/\,Keehnen, R.M.\,/\,Lindhout, E.\,/\,Zhou, D.F.\,/\,de 
Groot, C.\,/\,Pals, S.T.},
\normalsize
{\it Eur. J. Immunol.\/}
\normalsize
{\bf 27}
\normalsize
{\rm (1997)},
\normalsize
{\rm 1-7}.
\normalsize

\bibitem{Rad98}
\normalsize
{\sc Radmacher, M.D.\,/\,Kelsoe, G.\,/\,Kepler, T.B.},
\normalsize
{\rm Predicted and Inferred Waiting-Times for Key Mutations 
in the Germinal Center Reaction -- Evidence for Stochasticity 
in Selection}.
\normalsize
{\it Immunol. Cell Biol.\/}
\normalsize
{\bf 76}
\normalsize
{\rm (1998)},
\normalsize
{\rm 373-381}.
\normalsize

\bibitem{Pas94b}
\normalsize
{\sc Pascual, V.\,/\,Cha, S.\,/\,Gershwin, M.E.\,/\,Capra, J.D.\,/\,Leung, 
P.S.C.},
\normalsize
{\rm Nucleotide Sequence Analysis of Natural and Combinatorial 
Anti-PDC-E2 Antibodies in Patients with Primary Biliary 
Cirrhosis}.
\normalsize
{\it J. Immunol.\/}
\normalsize
{\bf 152}
\normalsize
{\rm 1994},
\normalsize
{\rm 2577-2585}.
\normalsize

\bibitem{Smi97}
\normalsize
{\sc Smith, K.\,/\,Light, A.\,/\,Nossal, G.\,/\,Tarlington, D.},
\normalsize
{\rm The extent of affinity maturation differs between the 
memory and antibody-forming cell compartments in the 
primary immune response}.
\normalsize
{\it EMBO J.\/}
\normalsize
{\bf 16}
\normalsize
{\rm (1997)},
\normalsize
{\rm 2996-3006}.
\normalsize

\bibitem{Cho98}
\normalsize
{\sc Choe, J.\,/\,Choi, Y.S.},
\normalsize
{\rm IL-10 Interrupts Memory B-Cell Expansion in the Germinal 
Center by Inducing Differentiation into Plasma-Cells}.
\normalsize
{\it Eur. J. Immunol.\/}
\normalsize
{\bf 28}
\normalsize
{\rm (1998)},
\normalsize
{\rm 508-515}.
\normalsize

\bibitem{Sie01}
\normalsize
{\sc Siepmann, K.\,/\,Skok, J.\,/\,van Essen, D.\,/\,Harnett, M.\,/\,Gray, 
D.},
\normalsize
{\rm Rewiring of CD40 is necessary for delivery of rescue 
signals to B cells in germinal centres and subsequent 
entry into the memory pool}.
\normalsize
{\it Immunol.\/}
\normalsize
{\bf 102}
\normalsize
{\rm (2001)},
\normalsize
{\rm 263-272}.
\normalsize

\bibitem{Cam98}
{\sc Camacho, S.A.\,/\,Kosco-Vilbois, M.H.\,/\,Berek, C.},
{\rm The Dynamic Structure of the Germinal Center}.
{\it Immunol. Today\/}
{\bf 19}
{\rm (1998)},
{\rm 511-514}.

\bibitem{Mey02}
\normalsize
{\sc Meyer-Hermann, M.},
\normalsize
{\rm A Mathematical Model for the Germinal Center Morphology 
and Affinity Maturation}.
\normalsize
{\it J. Theor. Biol.\/}
\normalsize
{\bf 216}
\normalsize
{\rm (2002)},
\normalsize
{\rm 273-300.}

\bibitem{Len01}
\normalsize
{\sc Lentz, V.M.\,/\,Manser, T.},
\normalsize
{\rm Germinal centers can be induced in the absence of T 
cells}.
\normalsize
{\it J. Immunol.\/}
\normalsize
{\bf 167}
\normalsize
{\rm (2001)},
\normalsize
{\rm 15-20}.
\normalsize

\bibitem{Kle01}
{\sc Kleinstein, S.H.\,/\,Singh, J.P.},
{\rm Toward quantitative simulation of germinal center dynamics:
Biological and modeling insights from experimental validation}.
{\it J. Theor. Biol.\/}
{\bf 211}
{\rm (2001)},
{\rm 253-275}.

\bibitem{Aga98}
\normalsize
{\sc Agarwal, A.\,/\,Nayak, B.P.\,/\,Rao, K.V.S.},
\normalsize
{\rm B-Cell Responses to a Peptide Epitope. VII. Antigen-Dependent 
Modulation of the Germinal Center Reaction}.
\normalsize
{\it J. Immunol.\/}
\normalsize
{\bf 161}
\normalsize
{\rm (1998)},
\normalsize
{\rm 5832-5841}.
\normalsize

\bibitem{Vin00}
\normalsize
{\sc de Vinuesa, C.G.\,/\,Cook, M.C.\,/\,Ball, J.\,/\,Drew, M.\,/\,Sunners, 
Y.\,/\,Cascalho, M.\,/\,Wabl, M.\,/\,Klaus, G.G.B.\,/\,MacLennan, C.M.},
\normalsize
{\rm Germinal centers without T cells}.
\normalsize
{\it J. Exp. Med.\/}
\normalsize
{\bf 191}
\normalsize
{\rm (2000)},
\normalsize
{\rm 485-493}.
\normalsize

\bibitem{Mey01}
\normalsize
{\sc Meyer-Hermann, M.},
\normalsize
{\rm Recycling Probability and Dynamical Properties of Germinal
Center Reactions}.
\normalsize
{\it J. Theor. Biol.\/}
\normalsize
{\bf 210}
\normalsize
{\rm (2001)},
\normalsize
{\rm 265-285}.
\normalsize

\bibitem{Per79}
\normalsize
{\sc Perelson, A.S.\,/\,Oster, G.F.},
\normalsize
{\rm Theoretical Studies of Clonal Selection: Minimal Antibody 
Repertoire Size and Reliability of Self-Non-self Discrimination}.
\normalsize
{\it J. Theor. Biol.\/}
\normalsize
{\bf 81}
\normalsize
{\rm (1979)},
\normalsize
{\rm 645-670}.
\normalsize

\bibitem{Wei70}
\normalsize
{\sc Weigert, M.\,/\,Cesari, I.\,/\,Yonkovitch, S.\,/\,Cohn, M.},
\normalsize
{\rm Variability in the light chain sequences of mouse antibody}.
\normalsize
{\it Nature\/}
\normalsize
{\bf 228}
\normalsize
{\rm (1970)},
\normalsize
{\rm 1045-1047}.
\normalsize

\bibitem{Kue93}
\normalsize
{\sc K\"uppers, R.\,/\,Zhao, M.\,/\,Hansmann, M.L.\,/\,Rajewsky, K.},
\normalsize
{\rm Tracing B Cell Development in Human Germinal Centers 
by Molecular Analysis of Single Cells Picked from Histological 
Sections}.
\normalsize
{\it EMBO J.\/}
\normalsize
{\bf 12}
\normalsize
{\rm (1993)},
\normalsize
{\rm 4955-4967}.
\normalsize

\bibitem{Wed97}
\normalsize
{\sc Wedemayer, G.J.\,/\,Patten, P.A.\,/\,Wang, L.H.\,/\,Schultz, 
P.G.\,/\,Stevens, R.C.},
\normalsize
{\rm Structural insights into the evolution of an antibody 
combining site}.
\normalsize
{\it Science\/}
\normalsize
{\bf 276}
\normalsize
{\rm (1997)},
\normalsize
{\rm 1665-1669}.
\normalsize

\bibitem{Han95}
\normalsize
{\sc Han, S.H.\,/\,Hathcock, K.\,/\,Zheng, B.\,/\,Kepler, T.B.\,/\,Hodes, 
R.\,/\,Kelsoe, G.},
\normalsize
{\rm Cellular Interaction in Germinal Centers: Roles of 
CD40-Ligand and B7-1 and B7-2 in Established Germinal 
Centers}.
\normalsize
{\it J. Immunol.\/}
\normalsize
{\bf 155}
\normalsize
{\rm 1995},
\normalsize
{\rm 556-567}.
\normalsize

\bibitem{Shi02}
{\sc Shih, T.-A. Y.\,/\,Meffre, E.\,/\,Roederer, M.\,/\,Nussenzweig,
C.},
{\rm Role of BCR affinity in T cell-dependent antibody responses in
vivo}.
{\it Nature Immunol.\/}
{\bf 3}
{\rm 2002},
{\rm 570-575}.

\bibitem{Cho96}
\normalsize
{\sc Choe, J.\,/\,Kim, H.S.\,/\,Zhang, X.\,/\,Armitage, R.J.\,/\,Choi, Y.S.},
\normalsize
{\rm Cellular and molecular factors that regulate the differentiation 
and apoptosis of germinal center B cells. Anti-Ig down-regulates 
Fas expression of CD40 ligand-stimulated germinal center 
B cells and inhibits Fas-mediated apoptosis}.
\normalsize
{\it J. Immunol.\/}
\normalsize
{\bf 157}
\normalsize
{\rm (1996)},
\normalsize
{\rm 1006-1016}.
\normalsize

\bibitem{Hol99}
\normalsize
{\sc Hollmann, C.\,/\,Gerdes, J.},
\normalsize
{\rm Follicular Dendritic Cells and T-Cells -- Nurses and 
Executioners in the Germinal Center Reaction}.
\normalsize
{\it Journal Of Pathology\/}
\normalsize
{\bf 189}
\normalsize
{\rm (1999)},
\normalsize
{\rm 147-149}.
\normalsize

\bibitem{Eij01}
\normalsize
{\sc van Eijk, M.\,/\,Medema, J.P.\,/\,de Groot, C.},
\normalsize
{\rm Cellular Fas-Associated Death Domain-Like IL-1-Converting 
Enzyme-Inhibitory Protein Protects Germinal Center 
B Cells from Apoptosis Durin Germinal Center Reactions}.
\normalsize
{\it J. Immunol.\/}
\normalsize
{\bf 166}
\normalsize
{\rm (2001)},
\normalsize
{\rm 6473-6476}.
\normalsize


\end{thebibliography}
\end{document}